\documentclass[epj]{webofc}
\usepackage[varg]{txfonts}   % Web of Conferences font
\woctitle{Physics Opportunities at an Electron-Ion Collider}

\def\empile#1\over#2{\mathrel{\mathop{\kern 0pt#1}\limits_{#2}}}

\newcommand{\slv}{\raise.15ex\hbox{$/$}\kern-.53em\hbox{$v$}}
\newcommand{\slF}{\raise.15ex\hbox{$/$}\kern-.53em\hbox{$F$}}
\newcommand{\slL}{\raise.15ex\hbox{$/$}\kern-.53em\hbox{$L$}}
\newcommand{\slP}{\raise.15ex\hbox{$/$}\kern-.53em\hbox{$P$}}
\newcommand{\slp}{\raise.15ex\hbox{$/$}\kern-.53em\hbox{$p$}}
\newcommand{\slq}{\raise.15ex\hbox{$/$}\kern-.53em\hbox{$q$}}
\newcommand{\slR}{\raise.15ex\hbox{$/$}\kern-.53em\hbox{$R$}}
\newcommand{\slQ}{\raise.15ex\hbox{$/$}\kern-.53em\hbox{$Q$}}
\newcommand{\slK}{\raise.15ex\hbox{$/$}\kern-.53em\hbox{$K$}}
\newcommand{\slk}{\raise.15ex\hbox{$/$}\kern-.53em\hbox{$k$}}
\newcommand{\slD}{\raise.15ex\hbox{$/$}\kern-.73em\hbox{$D$}}
\newcommand{\slC}{\raise.15ex\hbox{$/$}\kern-.53em\hbox{$C$}}
\newcommand{\slA}{\raise.15ex\hbox{$/$}\kern-.53em\hbox{$A$}}
\newcommand{\slSigma}{\raise.15ex\hbox{$/$}\kern-.53em\hbox{$\Sigma$}}
\newcommand{\slpartial}{\raise.15ex\hbox{$/$}\kern-.53em\hbox{$\partial$}}
\newcommand{\slcalP}{\raise.15ex\hbox{$/$}\kern-.63em\hbox{$\cal P$}}

\def\p{{\boldsymbol p}}

\def\kk{{\boldsymbol k}}

\begin{document}
\title{Kinetic theory of a longitudinally expanding system}

\author{Fran\c ois Gelis\inst{1}\fnsep\thanks{\email{francois.gelis@cea.fr}}
}

\institute{Institut de physique th\'eorique, CEA, CNRS, Universit\'e Paris-Saclay\\ F-91191 Gif-sur-Yvette, France}

\abstract{We use kinetic theory in order to study the role of quantum
  fluctuations in the isotropization of the pressure tensor in a
  system subject to fast longitudinal expansion, such as the matter
  produced in the early stages of a heavy ion collision.}
\maketitle
\section{Introduction}
\label{sec:intro}
Heavy ion collisions pose a number of interesting challenges to
Quantum Chromodynamics (QCD). Despite the very high center of mass
energy of these collisions, the transverse momentum of a typical final
state particle is rather small compared to the hard scales typical in
jet physics. Thus, a legitimate question is whether one can describe
the bulk of this particle production using a weak coupling approach in
QCD. The very high parton occupation numbers reached at these energies
are the key to a positive answer to this question. At high energy, the
gluon density in the hadron wavefunction increases exponentially with
rapidity, and eventually the gluons reach an occupation such that
their nonlinear interactions become important. These non-linearities
tame the growth of the gluon occupation number, and introduce a
dynamically generated dimensionful scale --the saturation momentum
$Q_s$-- in the problem.

\begin{figure}[htbp]
\centering
\includegraphics[width=12cm,clip]{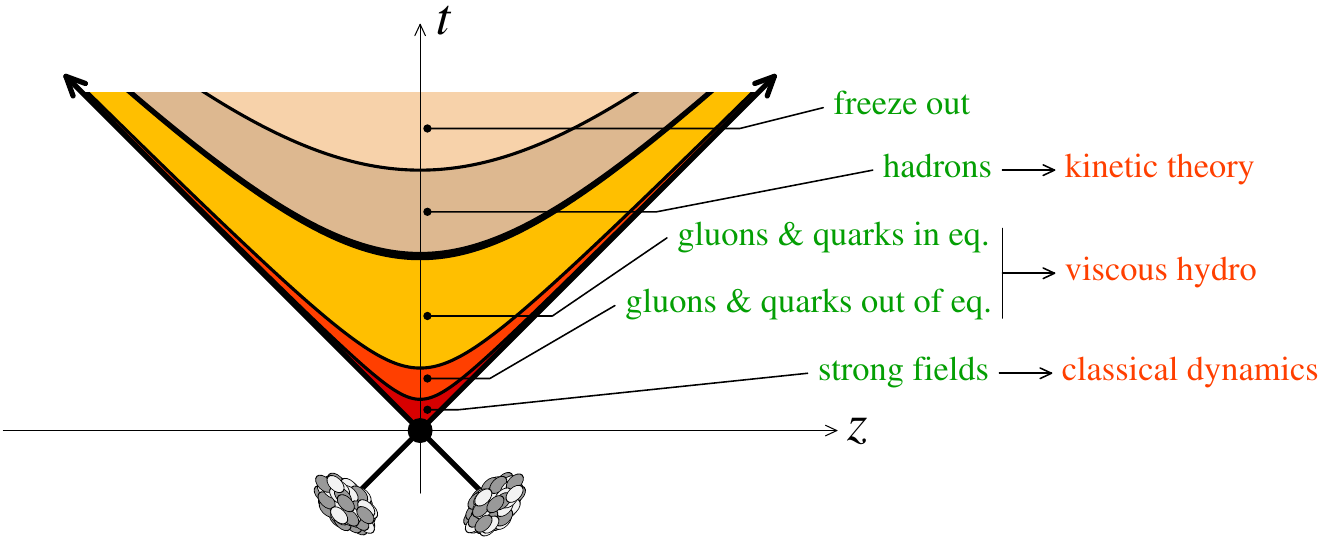}
\caption{Successive stages of a heavy ion collision.}
\label{fig:stages}
\end{figure}
An effective way of describing the physics of gluon saturation and
calculating QCD processes in this regime came a bit later, firstly in
the form of the McLerran-Venugopalan model \cite{McLerran:1994ni}
where the degrees of freedom were identified and the classical aspects
of their dynamics recognized. In this model, the degrees of freedom
with a large longitudinal momentum (in the observer's frame)
contribute via the color flow they carry, and are treated as static
variables thanks to Lorentz time dilation (the internal dynamics of a
hadron appears frozen in the observer's frame, on the timescale of the
collision). This simplification does not apply to gluons in the
vicinity of the observer's rapidity, and they are treated as
conventional gauge field operators.
\begin{figure}[htbp]
\centering
\includegraphics[width=12cm,clip]{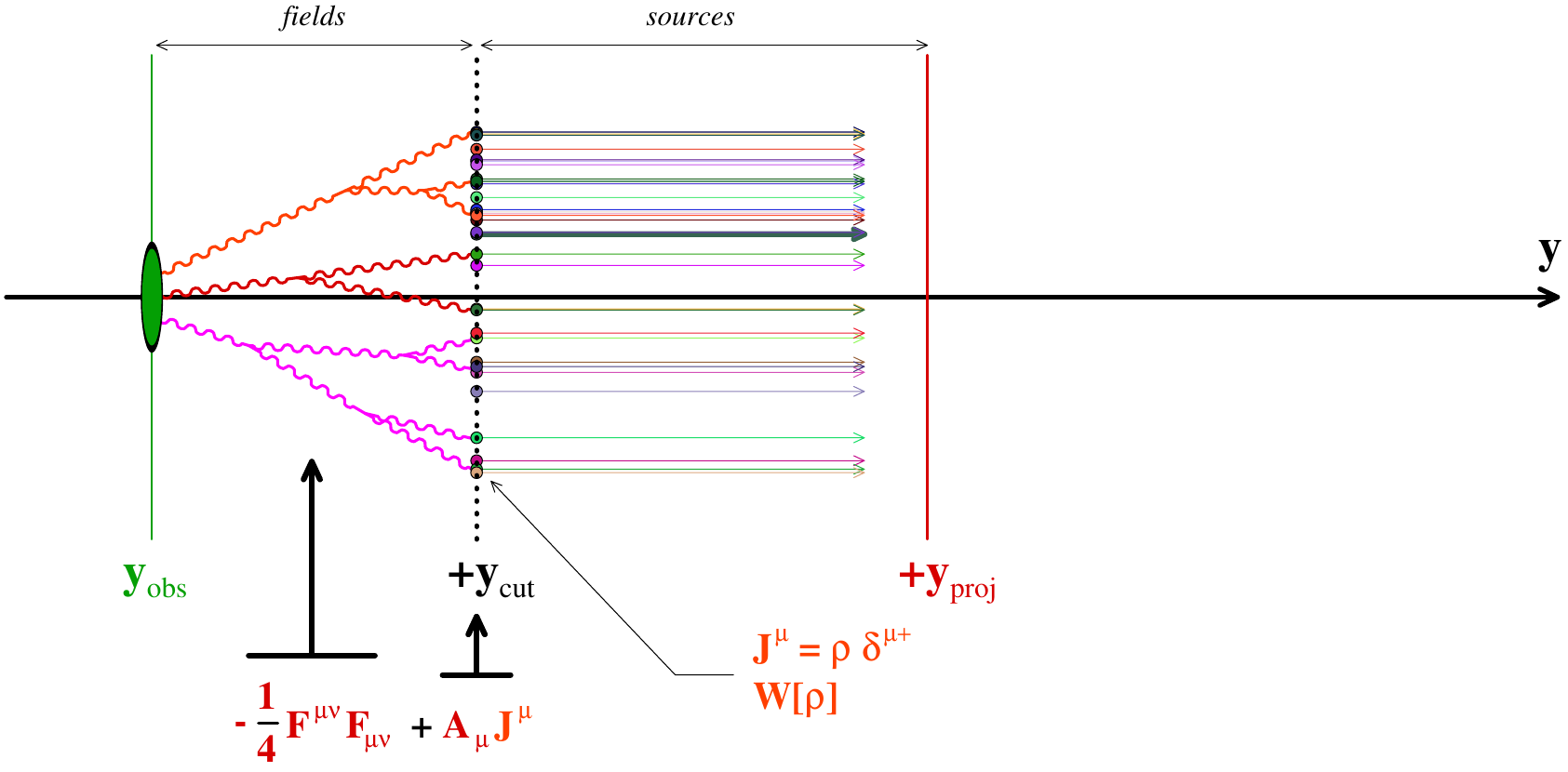}
\caption{Degrees of freedom in the CGC and their interplay.}
\label{fig:MV}
\end{figure}
In the MV model, the distribution of color sources in the wavefunction
of a high energy nucleus was argued to be Gaussian. A few years later,
it was realized that this distribution must depend on a longitudinal
momentum scale, due to large logarithms coming from loop corrections.
This led to the Color Glass Condensate (CGC) effective theory, and to
the JIMWLK evolution equation for the distribution of sources
\cite{Gelis:2010nm}.

\section{Evolution towards hydrodynamics in the CGC}
\label{eq:hydro}
The CGC provides a tool for systematically calculating observable
quantities in heavy ion collisions, such as the gluon spectrum, the
energy-momentum tensor, etc... One may think of using the
energy-momentum tensor calculated in the CGC framework as an initial
condition for hydrodynamical models of the expansion of the fireball
produced in heavy ion collisions. Such a matching is justified
theoretically provided that the two descriptions (CGC and
hydrodynamics) agree over some non-empty time range.

\begin{figure}[htbp]
\centering
%\sidecaption
\includegraphics[width=9cm,clip]{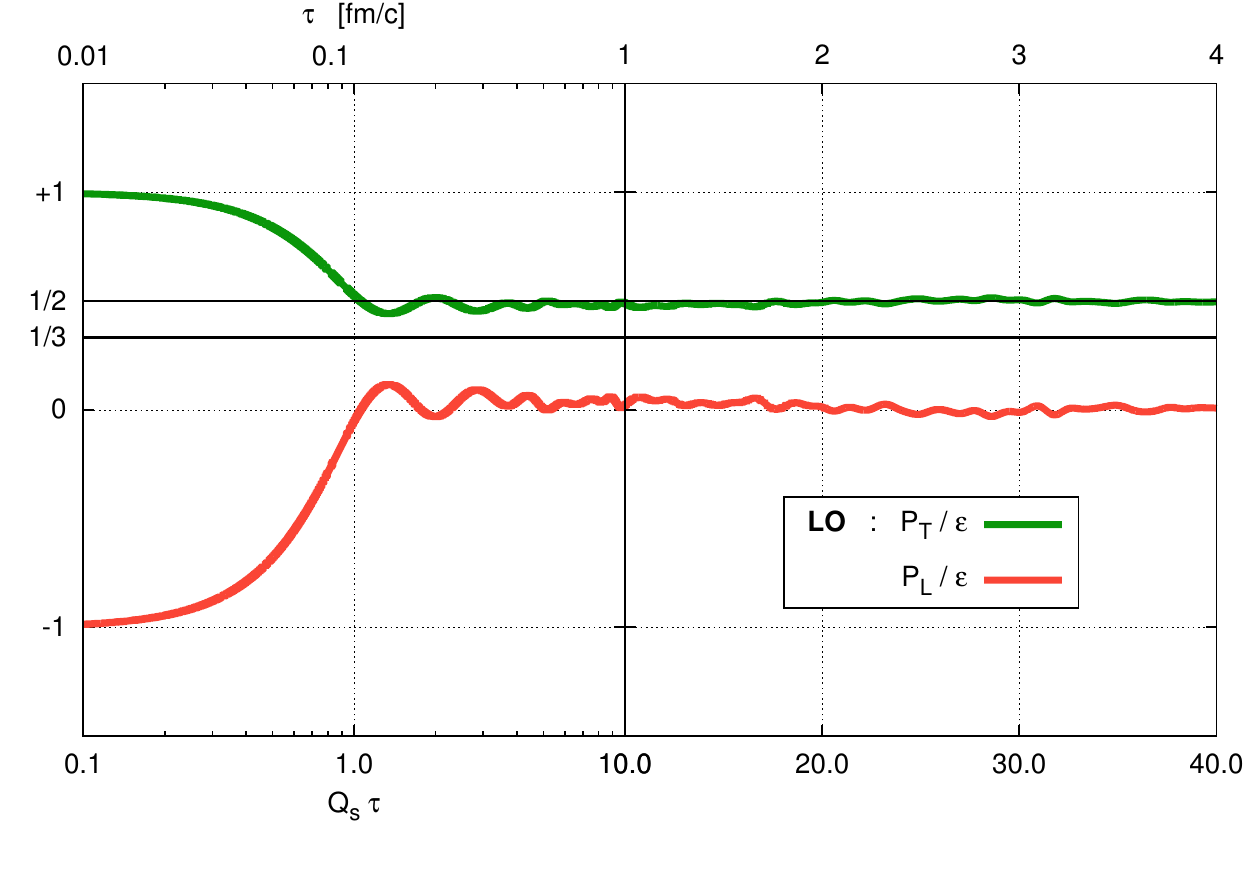}
\caption{Ratio of the transverse and longitudinal pressures to the
  energy density in the CGC at leading order. The lower times axis is
  in units of $1/Q_s$. The upper axis gives absolute values for the
  specific choice $Q_s=2~$GeV.}
\label{fig:LO}
\end{figure}
At leading order, the CGC leads to a very simple answer, since all
inclusive observables can be expressed in terms of a classical
solution of the Yang-Mills equations. However, this classical
approximation leads to an energy-momentum tensor whose behavior is
difficult to reconcile with hydrodynamics. Just after the collision,
the longitudinal pressure is negative.  It increases and becomes
mostly positive at a time $Q_s\tau\sim 1$. However, the ratio of
longitudinal to transverse pressure decreases forever, while it increases in
hydrodynamics.

In higher orders, the energy-momentum tensor receives quantum
contributions, beyond that of the LO classical fields. Despite being
suppressed by a power of $\alpha_s$, these corrections may be large
because of instabilities that exist in classical solutions of
Yang-Mills equations. Indeed, NLO corrections can be expressed in
terms of perturbations around the LO classical solutions, that grow
exponentially in the presence of unstable modes.

Thus, instead of a strict NLO calculation, one should seek for a
resummation. One such framework is provided by the \emph{classical
  statistical approximation} (CSA), in which one performs an average
over classical solutions obtained from Gaussian fluctuating initial
conditions. In the limit where the variance of the fluctuations goes
to zero, one recovers a unique classical evolution. Several
implementations of this scheme have been proposed in the literature,
that differ in the distribution of fluctuating initial fields.

{\bf 1.} There exists an ensemble of initial fields such that the CSA
contains the leading and next-to-leading orders exactly
\cite{Epelbaum:2013waa,Gelis:2013rba}, and a subset of all higher
orders. In order to achieve this, the initial fields must be the sum
of the leading order classical field ${\cal A}^{\mu a}(x)$ (non
fluctuating) and a linear superposition of \emph{mode functions}
$a^{\mu a}_{\kk\lambda c}(x)$:
\begin{eqnarray}
&&
A^{\mu a}(x)={\cal A}^{\mu a}(x)+ \sum_{\lambda,c}\int_\kk\frac{d^3\kk}{(2\pi)^3 2|\kk|} \; \Big[
c_{\kk\lambda c}\,a^{\mu a}_{\kk\lambda c}(x)+\mbox{c.c.}\Big]\; ,\nonumber\\
&&\big<c_{\kk\lambda c}\big>=0\;,\quad \big<c_{\kk\lambda c}^{\ }c_{\kk'\lambda' c'}^*\big>
=\frac{1}{2}\;2\big|\kk\big|\,(2\pi)^3\delta(\kk-\kk')\,\delta_{\lambda\lambda'}\,\delta_{cc'}\; .
\label{eq:RF1}
\end{eqnarray}
In this formula, the functions $a^{\mu a}_{\kk\lambda c}(x)$ are
solutions of the Yang-Mills equations linearized about the field
${\cal A}^{\mu a}$; with plane wave initial conditions (of momentum
$\kk$, polarization $\lambda$ and color $c$). The coefficients
$c_{\kk\lambda c}$ are random Gaussian distributed, with a null mean
value and the variance given in the second line. The prefactor $1/2$
in this variance can be interpreted as the occupation number
corresponding to the quantum fluctuations in each quantum mode.  Since
this setup reproduces the one loop result, it also leads to the usual
one-loop ultraviolet divergences. In addition, it also contains extra
ultraviolet divergences that do not exist in the underlying
theory. These spurious divergences originate from the fact that the
CSA is a non-renormalizable scheme: starting at two loops, it contains
divergences that cannot be removed by a renormalization of the
parameters of the Lagrangian \cite{Epelbaum:2014yja}. This leads to a
very strong dependence on the ultraviolet cutoff when it is chosen
much larger than the physical scales \cite{Berges:2013lsa,Epelbaum:2014mfa}.

{\bf 2.} Another implementation of the CSA uses as initial condition
fluctuating fields that correspond to a gas of gluons with a
distribution $f(\kk)$. These initial fields have no non-fluctuating
part, and the variance of the random coefficients is proportional to
$f(\kk)$ \cite{Berges:2013eia,Berges:2013fga,Berges:2014bba}:
\begin{eqnarray}
&&
A^{\mu a}(x)= \sum_{\lambda,c}\int_\kk\frac{d^3\kk}{(2\pi)^3 2|\kk|} \; \Big[
c_{\kk\lambda c}\,a^{\mu a}_{\kk\lambda c}(x)+\mbox{c.c.}\Big]\; ,\nonumber\\
&&\big<c_{\kk\lambda c}\big>=0\;,\quad \big<c_{\kk\lambda c}^{\ }c_{\kk'\lambda' c'}^*\big>
=f(\kk)\;2\big|\kk\big|\,(2\pi)^3\delta(\kk-\kk')\,\delta_{\lambda\lambda'}\,\delta_{cc'}\; .
\label{eq:RF2}
\end{eqnarray}
(Note the fact that in eq.~(\ref{eq:RF2}) the distribution $f(\kk)$
replaces the $1/2$ of eq.~(\ref{eq:RF1}).)  This implementation of the
CSA leads to ultraviolet finite results provided that the distribution
$f(\kk)$ decreases sufficiently fast, but it does not contain any
quantum fluctuations.

Eqs.~(\ref{eq:RF1}) and (\ref{eq:RF2}) describe rather different
situations, despite their resemblance: the flat spectrum of
Eq.~(\ref{eq:RF1}) on top of a non-fluctuating classical field
describes a quantum coherent state, while the spectrum of
Eq.~(\ref{eq:RF2}), with compact support, describes an incoherent
classical state.  When applied to simulations of the early stages of
heavy ion collisions, these two types of fluctuating initial
conditions lead to quite different behaviors of the pressure
tensor. The setup {\bf 1} leads to a roughly constant ratio
$P_{_L}/P_{_T}$ on short timescales, of the order of a few $Q_s^{-1}$,
while the setup {\bf 2} leads to a ratio $P_{_L}/P_{_T}$ that
decreases forever. Based on the time evolution of the particle
distribution in the second setup, it has been argued that quantum
corrections may be ignored up to large times of order
$\alpha_s^{-3/2}Q_s^{-1}$.

\section{Kinetic theory approach}
\label{sec:kin}
To clarify the role of quantum fluctuations in the process of
isotropization, one should include them while preserving
renormalizability.  A field theoretic framework that does this is the
\emph{2-particle irreducible} approximation
\cite{Luttinger:1960ua,Aarts:2002dj,Calzetta:2002ub,Berges:2004pu} of
the Kadanoff-Baym equations. Although in principle applicable to a
longitudinally expanding system \cite{Hatta:2011ky,Hatta:2012gq}, this
still remains a challenging task. A computationally much cheaper
alternative is to use kinetic theory \cite{Epelbaum:2014mfa} in order
to investigate the role of quantum fluctuations. Indeed, the
implementations {\bf 1} and {\bf 2} of the CSA described in the
previous section correspond to definite Feynman rules in which certain
vertices are omitted and where a propagator is approximated. By using
these truncated Feynman rules in the calculation of the appropriate
self-energy diagrams, one can obtain the kinetic analogue of the
approximations {\bf 1} and {\bf 2}. The Boltzmann equation with $2\to
2$ scatterings reads:
\begin{eqnarray}
\partial_t f_3
&\sim&
g^4\int_{124}
\cdots
\big[f_1f_2({f_3+f_4})-{f_3f_4}(f_1+f_2)\big]
\nonumber\\
&&\quad+g^4\int_{124}\cdots\big[f_1f_2-{f_3f_4}\big]\; .
\label{eq:boltz}
\end{eqnarray}
where the dots contain the cross-section and the delta functions for
the conservation of energy and momentum.  In Eq.~(\ref{eq:boltz}), the
first line contains the terms that remain in the classical
approximation of Eq.~(\ref{eq:RF2}) (see
refs.~\cite{Mueller:2002gd,Jeon:2004dh}), and the terms on the second
line come from quantum fluctuations. Ignoring these terms quadratic in
the distribution $f$ is the kinetic theory analogue of the
implementation {\bf 2} of the CSA.

\begin{figure}[htbp]
\centering
%\sidecaption
\includegraphics[width=4cm,clip]{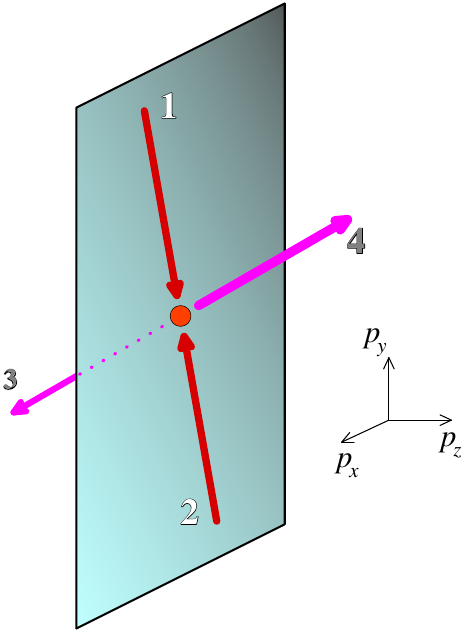}
\caption{Kinematics of a $2\to 2$ scattering process that would
  contribute to the isotropization of pressures when the particle
  distribution is very anisotropic (with $p_z\approx 0$). The momenta
  of the two particles in the initial state are purely transverse, and
  the outgoing particles have opposite $p_z$'s.}
\label{fig:iso}
\end{figure}
The cubic terms of the first line are dominant in the regime of large
occupation number. But since this approximation is usually not uniform
across all momentum space, this may lead to problems especially when
the distribution is very anisotropic.  In Fig.~\ref{fig:iso}, we
illustrate this for such a distribution, for which the incoming
particles 1,2 have almost purely transverse momenta. Isotropization
would require that one of the outgoing particles (3 or 4) has a
nonzero $p_z$. With this kinematics, non vanishing contributions can
only come from the second line, which is neglected in the classical
approximation.

\begin{figure}[htbp]
\centering
%\sidecaption
\includegraphics[width=7cm,clip]{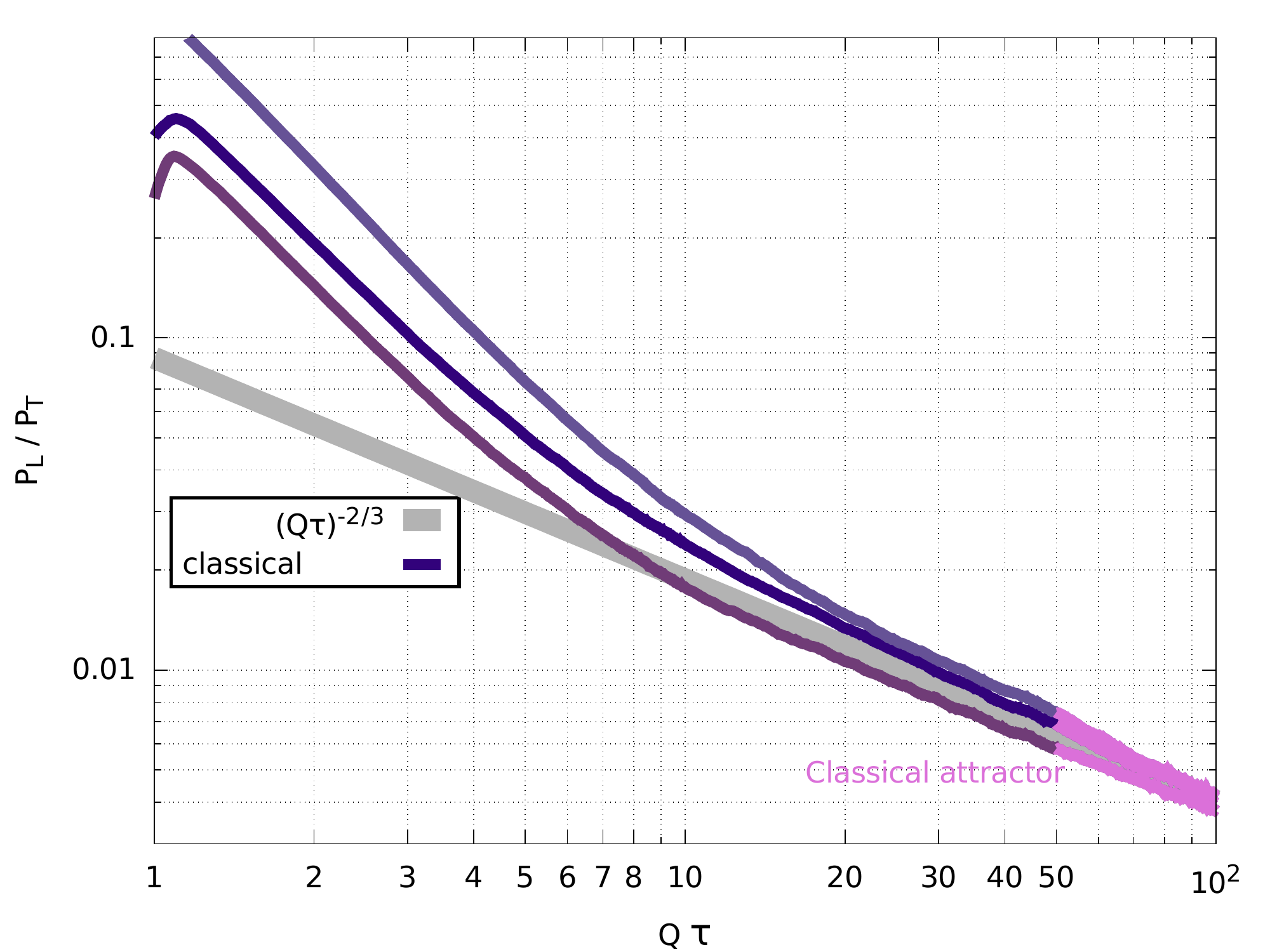}
\caption{Time evolution of the ratio $P_{_L}/P_{_T}$ in the classical
  approximation of the Boltzmann equation. When the initial condition
  is of the form $f_0(\p)=(n_0/g^2){\bf f}(\p)$, the classical
  evolution does not depend on the coupling and its value need not be
  specified. The three curves correspond to various CGC-like initial
  conditions that differ in the magnitude of the initial
  anisotropy. All these evolutions converge towards a universal
  classical attractor.}
\label{fig:class}
\end{figure}

\begin{figure}[htbp]
\centering
%\sidecaption
\includegraphics[width=7cm,clip]{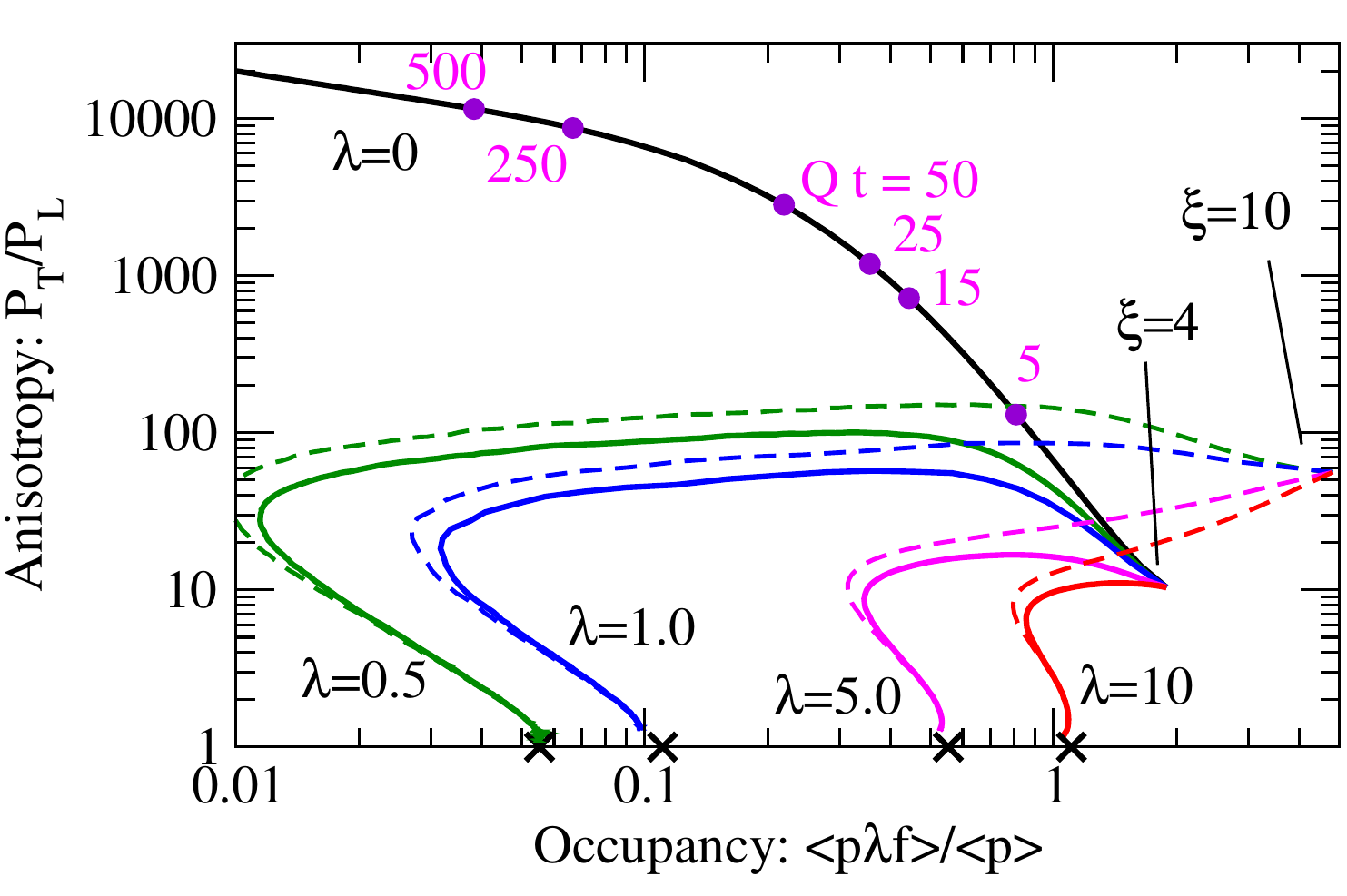}
\caption{Evolution of the anisotropy and momentum weighted occupation
  number in a system of gluons described in kinetic theory, for
  initial conditions of the form $f_0(\p)=(n_0/\lambda){\bf
    f}(\p)$. The curve labeled $\lambda=0$ is a classical
  approximation, while the curves for $\lambda=0.5, 1, 5, 10$ are
  obtained with the full Boltzmann equation. Plot taken from
  \cite{Kurkela:2015qoa}, with time labels added by us for clarity.}
\label{fig:AK}
\end{figure}

This has been seen in numerical studies of the Boltzmann equation for
a longitudinally expanding system, both for scalar fields
\cite{Epelbaum:2015vxa} and for gluons \cite{Kurkela:2015qoa}, with
CGC-like highly occupied initial distributions of the form
$f_0(\kk)=(n_0/g^2)\,\theta(Q_s-k_\perp)\,\theta(\xi Q_s-|k_z|)$ at a
time of order $Q_s^{-1}$. When one keeps only the classical terms, the
Boltzmann equation leads to a forever decreasing $P_{_L}/P_{_T}$ (see
Fig.~\ref{fig:class}, and the curve $\lambda=0$ in
Fig.~\ref{fig:AK}). Moreover, like in the setup {\bf 2} of the CSA,
the classical approximation of the Boltzmann equation leads to a
\emph{classical attractor}: the asymptotic behavior of these classical
evolutions is universal, regardless of the details of the initial
condition.

\begin{figure}[htbp]
\centering
%\sidecaption
\includegraphics[width=7cm,clip]{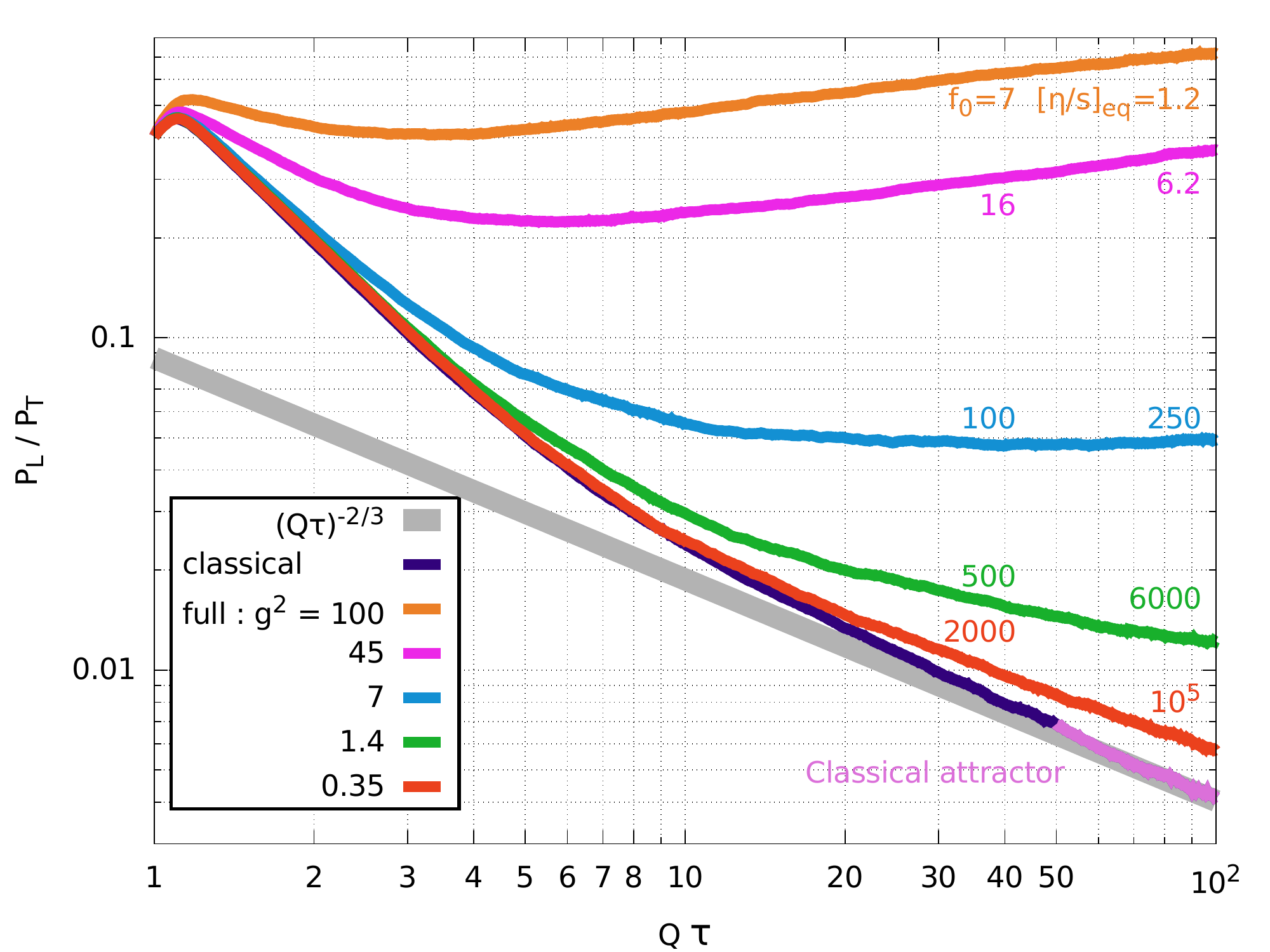}
\caption{Time evolution of the ratio $P_{_L}/P_{_T}$ obtained from the
  full Boltzmann equation, for various values of the coupling
  constant. The initial condition is of the form
  $f_0(\p)=(n_0/g^2){\bf f}(\p)$. The evolution in the classical
  approximation is shown for the same initial condition for
  comparison. The numbers $f_0$ and $[\eta/s]_{\rm eq}$ overlaid on
  the plot indicate the corresponding initial occupation number and
  the equilibrium value of the viscosity to entropy ratio (in
  equilibrium).}
\label{fig:class+quant}
\end{figure}

The figure \ref{fig:class+quant} (for scalar fields) and the curves
labeled $\lambda=0.5$ to $10$ in Fig.~\ref{fig:iso} (for gluons in
Yang-Mills theory) show the results obtained with the full Boltzmann
equation, including the terms of quantum origin (the second line in
eq.~(\ref{eq:boltz})).  These results show that it takes very small
couplings (or equivalently, very large values of the ratio $\eta/s$)
and very large occupation numbers for the quantum evolution to get
close to the universal classical behavior. For larger yet still small
couplings, the classical and quantum evolutions depart from each other
very early: for instance, in the Yang-Mills case at $\lambda=0.5$
(i.e. $\alpha_s\approx 0.02$ for $N_c=2$ colors), this happens around
$Q_s\tau\approx 2$, well before the conjectured limit of validity of
the classical approximation, $Q_s\tau\approx \alpha_s^{-3/2}\approx
350$. This indicates that quantum fluctuations are essential for
isotropization: purely classical approximations do not correctly
capture the relevant physics and largely overestimate their own range
of validity.

\section*{Acknowledgements}
This work is supported by the Agence
Nationale de la Recherche project 11-BS04-015-01.

% BibTeX or Biber users please use (the style is already called in the class, ensure that the "woc.bst" style is in your local directory)

%\bibliography{spires}
%\bibliographystyle{woc}

\end{document}